\documentstyle[amssymb,preprint,aps,prl]{revtex}

\newcommand{\ud}{\text{d}}
\newcommand{\ui}{\text{i}}
\newcommand{\ue}{\text{e}}
\newcommand{\uTr}{\text{Tr}}

\newcommand{\R}{{\mathbb{R}}}
\newcommand{\C}{{\mathbb{C}}}

\newcommand{\vecA}{\bbox{A}}
\newcommand{\vecB}{\bbox{B}}
\newcommand{\vecE}{\bbox{E}}
\newcommand{\vecx}{\bbox{x}}
\newcommand{\vecy}{\bbox{y}}
\newcommand{\vecp}{\bbox{p}}
\newcommand{\vecs}{\bbox{s}}
\newcommand{\vecsig}{\bbox{\sigma}}
\newcommand{\vecalph}{\bbox{\alpha}}
\newcommand{\vecpi}{\bbox{\pi}}
\newcommand{\vecnab}{\bbox{\nabla}}
\newcommand{\vecxi}{\bbox{\xi}}

\begin{document}

\preprint{ULM-TP/98-3}

\title{Semiclassical Time Evolution and Trace Formula for Relativistic
Spin-1/2 Particles}

\author{Jens Bolte\footnote[1]{E-mail address: bol@physik.uni-ulm.de} \ and 
        Stefan Keppeler\footnote[3]{E-mail address: kep@physik.uni-ulm.de}\\}
\address{
Abteilung Theoretische Physik, Universit\"at Ulm\\
Albert-Einstein-Allee 11, D-89069 Ulm\\
Germany
}

\date{May 14, 1998}
\maketitle

\vfill

\begin{abstract}
We investigate the Dirac equation in the semiclassical limit $\hbar \to 0$. 
A semiclassical propagator and a trace formula are derived and are shown to be
determined by the classical orbits of a relativistic point particle. In 
addition, two phase factors enter, one of which can be calculated from the 
Thomas precession of a classical spin transported along the particle orbits. 
For the second factor we provide an interpretation in terms of dynamical and 
geometric phases.
\end{abstract}


The first one to seek a semiclassical treatment of the Dirac equation in the
spirit of the WKB-method appears to be Pauli \cite{Pau32}, who gave a solution 
for a special case. He found that in the semiclassical limit the translational 
motion is independent of the spin degrees of freedom. Because of this fact 
the formalism was criticized by de~Broglie \cite{Bro52} with the remark that 
one would expect `classical objects' like electromagnetic moments to influence 
the trajectories. This controversy was clarified by Rubinow and Keller 
\cite{RubKel63} in a paper that seems to have been overlooked by some later 
authors. Rubinow and Keller pointed out that the moments of an electron are 
proportional to $\hbar$ so that in leading order as $\hbar\to 0$ the influence 
of spin on the trajectories vanishes. However, in next-to-leading order the 
dynamical equation for Thomas precession \cite{Tho27} is obtained from the 
Dirac equation. Since only the ratio of the magnetic moment and spin enters 
this equation, it contains no $\hbar$ and therefore can be interpreted as 
describing the dynamics of a classical spin.

The general set-up for semiclassical quantization in the case of multicomponent
wave equations was developed by Littlejohn and Flynn \cite{LitFly91}
. In a short-wavelength approximation they replaced the matrix-valued wave 
operator
by a matrix-valued Hamiltonian function, such that its eigenvalues generate 
Hamiltonian dynamics in phase space. But even if these are integrable, an 
application of EBK quantization was found to be obstructed by the presence of 
additional phases. In \cite{LitFly91} a formalism was presented that allows 
to treat matrix Hamiltonians with no (globally) degenerate eigenvalues. 
Subsequently, Emmrich and Weinstein \cite{EmmWei96} outlined how to proceed in
the degenerate case that, e.~g., occurs for the Dirac equation, and pointed out
the problems of formulating a Bohr-Sommerfeld quantization. They, moreover, 
uncovered the global geometric meaning of the additional phases. A 
semiclassical quantization for special configurations, based on the complex
WKB-method, is presented in \cite{BagBelTriYev94}. 

In this paper we will follow an alternative approach in that we investigate the
semiclassical time evolution and then set up a trace formula. This
procedure avoids (some) difficulties that one encounters with semiclassical
approximations to eigenspinors and, furthermore, is not restricted to 
classically
integrable systems. We basically follow the approach that was developed by 
Gutzwiller \cite{Gut90} for the Schr\"odinger equation. Hence the basic object 
to be studied is the integral kernel $K(\vecx,\vecy,t)$ of the time evolution 
operator $U(t)$. Gutzwiller represented the kernel by a path integral and 
evaluated this semiclassically. However, here we prefer to use a representation
of the kernel in terms of an oscillatory integral. This procedure can be made 
mathematically rigorous as, e.~g., explained in \cite{Rob87} for the 
Schr\"odinger equation. In a second step we pass to the energy domain via 
Fourier transform, and then take the trace over spatial coordinates as well
as over spin degrees of freedom. This results in a periodic orbit formula for 
spectral functions. Special attention is paid to the r\^ole of spin. 
Our philosophy of a systematic semiclassical expansion in the context of the 
Dirac equation automatically ensures that spin is treated quantum mechanically 
from the outset, without any ad hoc semiclassical approximation. As mentioned 
above, the semiclassical asymptotics introduces an adiabatic decoupling of 
(slow) translational and (fast) spin degrees of freedom. This happens in such 
a way that to lowest orders in $\hbar$ the expected dynamical equations for 
both 
kinds of degrees of freedom emerge. In addition, our procedure allows to 
re-interpret the additional phases in terms of dynamical and geometric phases
associated with a precessing spin. The degree of freedom that is lost upon 
passing from a quantum mechanical description of spin in terms of 
SU(2)-matrices to a classical description in terms of vectors ${\bf s}\in\R^3$ 
with fixed length $|{\bf s}|$ can be reconstructed from one of these phases.
A detailed account of our approach will be presented elsewhere \cite{BolKep98}.

Let us now briefly summarize the calculations and results. We investigate the 
Dirac equation 
\begin{equation}
\label{Diraceqn}
  \ui\hbar\,\frac{\partial\Psi(\vecx,t)}{\partial t} = H_D \Psi(\vecx,t) 
\end{equation}
with the (quantum) Hamiltonian
\begin{equation}
\label{DiracHam} 
  H_D := \vecalph \left( \frac{\hbar}{\ui}\vecnab - 
        \frac{e}{c} \vecA (\vecx) \right) 
        + \beta mc^2 + e \, \varphi(\vecx) 
\end{equation}
that acts on a suitable domain in the Hilbert space $L^2(\R^3)\otimes\C^4$. The 
Dirac algebra is realized by 
\begin{equation}
  \bbox{\alpha} = \left( \begin{array}{cc}  0 & \bbox{\sigma} \\ 
                                            \bbox{\sigma} & 0 
                 \end{array} \right)
  \quad \text{and} \quad
  \beta = \left( \begin{array}{cc}  \openone_{2 \times 2} & 0 \\
                                    0 & - \openone_{2 \times 2}   
          \end{array} \right)\ ,
\end{equation}
where $\bbox{\sigma}$ is the vector of Pauli matrices. The time evolution 
kernel is defined by
\begin{equation}        
  \Psi({\bf x},t) = 
  \int_{\R^3} K({\bf x},{\bf y},t) \, \Psi_0({\bf y}) \ \ud^3 y
\end{equation}
so that it has to fulfill the Dirac equation for $t>0$ with initial 
condition $K(\vecx,\vecy,0) = \openone_{4 \times 4} \delta(\vecx-\vecy)$.
Anticipating the occurrence of solutions of appropriate classical equations of 
motion for positive and negative energies, respectively, we choose the 
semiclassical ansatz  
\begin{equation}
\label{scansatz}
  K(\vecx,\vecy,t) = \frac{1}{(2 \pi \hbar)^3} \int_{\R^3} \left[ 
  a^+_{\hbar} \ue^{\frac{\ui}{\hbar} \phi^+} 
  + a^-_{\hbar} \ue^{\frac{\ui}{\hbar} \phi^-}
  \right] \ud^3 \xi
\end{equation} 
with phase functions $\phi^{\pm}=\phi^{\pm} (\vecx,\vecy,t;\vecxi)$. The
amplitudes $a^\pm_{\hbar}$ are $4\times 4$--matrices with semiclassical
expansions
\begin{equation}
\label{scamplitude}
  a^{\pm}_{\hbar}(\vecx,\vecy,t;\vecxi) = \sum_{k=0}^{\infty} 
  (-\ui\hbar)^k \, a^{\pm}_k(\vecx,\vecy,t;\vecxi)\ .
\end{equation}
In order to account for the initial condition of the kernel, we have to choose
$\phi^{\pm}|_{t=0}=(\vecx -\vecy)\vecxi$ and $a^+_{\hbar}|_{t=0} + a^-_{\hbar
}|_{t=0} =\openone_{4 \times 4}$. Inserting (\ref{scansatz}) into (\ref{Diraceqn}) 
and comparing like orders in $\hbar$ yields to lowest order matrix equations which 
have solutions with non-zero $a^{\pm}_0$ only if $\phi^{\pm}$ satisfy the 
Hamilton-Jacobi equations 
\begin{equation}\label{HJG}
  H^{\pm}(\vecnab_{\vecx} \phi^{\pm},\vecx) 
  + \frac{\partial \phi^{\pm}}{\partial t} = 0
\end{equation}
with the (classical) Hamiltonians 
\begin{equation}
  H^{\pm}(\vecp,\vecx) = 
  e\varphi(\vecx)\pm\sqrt{c^2\left(\vecp -\frac{e}{c}\vecA(\vecx)\right)^2 
  + m^2c^4}\ .
\end{equation}
These are the (twofold degenerate) eigenvalues of the matrix-valued symbol 
$\vecalph(\vecp-e/c\vecA)+\beta mc^2+e\varphi$ of $H_D$. Due to (\ref{HJG}) one
can separate $\vecy$ in $\phi^{\pm}$ according to 
$\phi^{\pm}=S^{\pm}(\vecx,\vecxi,t)-\vecy\vecxi$. When one applies the 
method of stationary phase to (\ref{scansatz}) as $\hbar \to 0$, it turns out 
that at stationary points $S^{\pm}$ generates a canonical transformation that 
describes the dynamics of a relativistic point particle from $\vecy$ to $\vecx$
in time $t$. 

We now turn to the equations that occur in next-to-leading order in $\hbar$, 
and which contain terms involving both $a_0^\pm$ and $a_1^\pm$. 
Here we restrict to the index $+$. An equation for $a_0^+$ only is obtained 
through a multiplication on the left with the hermitian conjugate $V_t^{\dag}$ 
of the $4\times 2$--matrix
\begin{equation}
\label{Vdef}  
  V_t := V(\vecx,\vecy,t;\vecxi) = \frac{1}{\sqrt{2 \epsilon (\epsilon + mc^2)}}
      \left( \begin{array}{c} \epsilon + mc^2 \\ \vecsig \vecpi 
             \end{array} \right)
\end{equation}
with $\epsilon :=\sqrt{c^2\vecpi^2 +mc^2}$ and $\vecpi =\vecnab_{\vecx}\phi - 
\frac{e}{c} \vecA$, whose columns are the eigenvectors associated with $H^+$.
(We will denote by $W_t$ the corresponding matrix of eigenvectors associated 
with $H^-$.) We then define a $2\times 2$--matrix $b_+$ by
\begin{equation}\label{a0ansatz}
  a_0^+ = V_t b_+ V_0^{\dag}
\end{equation}
and remark that only this construction with $V_t$ on the left, together with 
the 
appropriate $\phi^+$, ensures that the equation to {\it lowest} order in 
$\hbar$ is fulfilled. Moreover, at $t=0$ the initial condition 
$b_+|_{t=0}=\openone_{2 \times 2}$ ensures that $a_0^+|_{t=0}=V_0 V_0^{\dag}$ 
is the projector on the
$H^+$-eigenspace. A respective remark applies to $a_0^-$ so that the initial 
condition for $a_0^+ +a_0^-$ is fulfilled. An obvious interpretation of this 
ansatz is as follows. Given an initial 4-spinor $\Psi_0$ at time $t=0$, 
$V_0^{\dag}$ projects it onto the $H^+$-eigenspace and converts it to a 
2-spinor.
This is propagated to time $t$ and then $V_t$ maps the 2-spinor back to 
the 4-spinor representation. Using the ansatz (\ref{a0ansatz}) in the equation 
of next-to-leading order in $\hbar$ then yields the following transport equation 
for $b=b_+$,
\begin{equation}
\label{transportb}
  \left[ \vecnab_{\vecp} H^+(\vecnab_{\vecx} S^+, \vecx) \vecnab_{\vecx} 
         + \frac{\partial}{\partial t} \right] b =
  -(M_1 + \ui M_2)\, b \ ,
\end{equation}
with the hermitian $2\times 2$--matrices 
\begin{eqnarray}
\label{Mjdef} 
  &&
  M_1 := \frac{1}{2} \sum_{j=1}^3 \left( \sum_{k=1}^3
  \frac{\partial^2 H^+}{\partial p_j \partial p_k} 
  \frac{\partial^2 S^+}{\partial x_k \partial x_j} 
  + \frac{\partial^2 H^+}{\partial p_j \partial x_j} \right) \, 
  \openone_{2 \times 2} \ , 
  \nonumber\\ &&
  M_2 := - \frac{ec}{2 \epsilon} \vecsig \vecB  
  + \frac{ec^2}{2 \epsilon (\epsilon + mc^2)} \vecsig (\vecpi \times \vecE)\ . 
\end{eqnarray}     
We need to solve (\ref{transportb}) only along the orbits in phase space, in 
which case the left-hand side can be viewed as the total time derivative
$\dot{b}$ along these orbits. To arrive at
(\ref{transportb}) we used Coulomb gauge, but this doesn't restrict the result 
because it only contains the fields $\vecE$ and $\vecB$. The contribution 
to (\ref{transportb}) coming from $M_1$ is well known from the Schr\"odinger 
case \cite{Rob87}, and therefore the ansatz
\begin{equation}
  b = \sqrt{ \det \left(\frac{\partial^2 S^+}{\partial x_j \partial\xi_k}
      \right)}\ d\ ,
\end{equation}
with some $2\times 2$--matrix $d$, proves useful. From (\ref{transportb}) one 
then obtains the transport equation 
\begin{equation} \label{transportd}
  \dot{d} + \ui M_2\,d = 0\ ,\quad d|_{t=0}=\openone_{2\times 2}\ ,
\end{equation}
for $d$, which only involves the spin degrees of freedom. Due to the unitarity 
of the time evolution and the initial condition, $d$ has to be an SU(2)-matrix.

The additional phases discussed in \cite{LitFly91,EmmWei96} are caused by 
$M_2$.
The second term in $M_2$, see (\ref{Mjdef}), can be shown to be 
$V^\dag(\vecnab_{\vecp} H^+\vecnab_{\vecx}+ \frac{\partial}{\partial t})V$, 
and thus
is a projection of the natural connection on the trivial $\C^4$-bundle over
phase space onto the $H^+$-eigenbundle. According to \cite{Sim83,EmmWei96}, it 
hence is
the Berry term identified in \cite{LitFly91}. We call this SU(2)-Berry term
in order to distinguish it from the U(1)-phase originally introduced by Berry
\cite{Ber84}. The first term in $M_2$ then is the ``no name term'' of
\cite{LitFly91} that has been shown to be related to a Poisson curvature in 
\cite{EmmWei96}.
In fact, it measures to what extent the classical time evolution tends to leave
the $H^+$-eigenspace. In physical terms, the first (curvature) term is the 
interaction of spin and magnetic field, and the second (SU(2)-Berry) term 
represents the spin-orbit coupling. Analogous considerations apply 
to $H^-$.

We are now in a position to state the following semiclassical expression for 
the time evolution kernel,
\begin{eqnarray} \label{Ksc}
  && K(\vecx,\vecy,t) = 
  \nonumber\\ && \hspace{9 mm}
  \frac{1}{(2 \pi \ui \hbar)^{3/2}} 
  \Bigg[ 
  \sum_{\gamma^+_{xy}} V_t d_+V_0^{\dag} \, D_{\gamma_{xy}^+}^+ 
  \ue^{\frac{\ui}{\hbar} R^+_{\gamma^+_{xy}} 
    - \ui \frac{\pi}{2} \nu_{\gamma^+_{xy}}}
  + \sum_{\gamma^-_{xy}} W_t d_-W_0^{\dag} \, D_{\gamma_{xy}^-}^- 
  \ue^{\frac{\ui}{\hbar} R^-_{\gamma^-_{xy}} 
    - \ui \frac{\pi}{2} \nu_{\gamma^-_{xy}}}
  \Bigg] 
  \left\{ 1 + {\cal O} (\hbar) \right\}
  \nonumber\\ &&
  \text{with} \ 
  D_{\gamma_{xy}^{\pm}}^{\pm} = \sqrt{ \left| \det \left( 
  -\frac{\partial^2 R^{\pm}_{\gamma^{\pm}_{xy}}}{\partial x_j \partial 
  y_k} \right) \right| }\ ,
\end{eqnarray}
where $\gamma_{xy}^{\pm}$ labels the classical orbits that connect 
$\vecy$ and $\vecx$ in time $t$. $R^{\pm}$ is Hamilton's principal function, 
which is the Legendre transform of $S^{\pm}$ with respect to $\vecxi$, and 
$\nu^{\pm}$ is the Morse index of the corresponding orbit.

We are still left with the calculation of $d$. Since $d\in\text{SU}(2)$, we 
can use the representation 
\begin{equation} 
  d = \left( \begin{array}{cr} u & -\bar{v} \\ v & \bar{u} \end{array} \right)
  \quad \text{with} \quad |u|^2 + |v|^2 = 1\ .
\end{equation}
A candidate for a `classical spin' should be a vector $\vecs\in\R^3$ with fixed
length, which we find convenient to choose as $|\vecs|=1$. We thus seek a map 
$d\mapsto\vecs$ from SU(2) to 
$\text{S}^2\subset\R^3$. To achieve this we propose to use the well known Hopf 
map $\pi_H :\text{SU(2)}\to\text{S}^2$ defined by
\begin{equation}
\label{spinvector}
   \pi_H(d)=\vecs := \left( \begin{array}{c} 2\,\text{Re}(u \bar{v}) \\
                                           2\,\text{Im}(u \bar{v}) \\
                                           |u|^2 - |v|^2 
                   \end{array} \right) = (\bar{u},\bar{v})\ \vecsig 
                   \left( \begin{array}{c} u \\ v \end{array} \right)\ .
\end{equation}
The last equality reveals that $\vecs$ is also connected to a suitable spin 
expectation value. From (\ref{spinvector}) and (\ref{transportd}) it follows 
that $\vecs$ fulfills the classical equation
\begin{equation} \label{ThomasPre}
  \dot{\vecs} = \vecs \times \left( \frac{ec}{\epsilon} \vecB - 
  \frac{ec^2}{\epsilon (\epsilon + mc^2)} \vecpi \times \vecE \right)  
  \, , \
  \vecs|_{t=0} = \left( \begin{array}{c} 0 \\ 0 \\ 1 \end{array} \right)
  \, , 
\end{equation}  
describing a precessing spin. After \cite{Tho27}, this is commonly called 
Thomas precession, see also \cite{BarMicTel59}. Due to the initial condition, 
$\vecs\in\text{S}^2$ will stay
on the northern hemisphere for sufficiently small times. We then choose polar 
coordinates,
\begin{equation}
  \vecs = \left( \begin{array}{c} \sin\theta \, \cos\phi \\
                                  \sin\theta \, \sin\phi \\
                                  \cos\theta             \end{array} \right)\ ,
\end{equation}
which allows to calculate $d$ up to a phase $\eta$, where 
$u/|u|=e^{\ui\eta}$,
\begin{equation}\label{d(s)}
  d = \left( \begin{array}{cc} \cos(\theta/2)\,\ue^{\ui \eta} &
                               -\sin(\theta/2)\,\ue^{-\ui (\eta - \phi)} \\
                               \sin(\theta/2)\,\ue^{\ui (\eta - \phi)} &
                               \cos(\theta/2)\,\ue^{-\ui \eta}
      \end{array} \right) \, .
\end{equation}
The equation for $\eta$, which is obtained upon inserting (\ref{d(s)}) into 
(\ref{transportd}) and multiplying by $d^{\dag}$, can immediately be integrated,
\begin{eqnarray}
\label{quantal-phase}
  \eta = \frac{1}{2} \int_0^t \vecs \left( \frac{ec}{\epsilon} \vecB - 
     \frac{ec^2}{\epsilon (\epsilon + mc^2)} \vecpi \times \vecE \right) \ud t'
     + \frac{1}{2} \int_0^t (1 - \cos\theta) \, \dot{\phi} \, \ud t'\ .
\end{eqnarray}
The first term is a dynamical phase associated with the energy of a (classical)
magnetic moment in given electromagnetic fields, whereas the second term is a
geometric phase. We remark that
once $\vecs$ enters the southern hemisphere of $\text{S}^2$, one should change
the phase convention in that $v/|v|=e^{\ui\lambda}$ is used to describe the 
non-classical degree of freedom. In (\ref{quantal-phase}) this amounts to 
replace $1/2(1 - \cos\theta)\ud\phi$ by $-1/2(1 +\cos\theta)\ud\phi$. These
two expressions are the well known gauges of the vector potential for a 
magnetic monopole of strength $1/2$ situated at the origin of the sphere.  
The geometric phase caused by this connection is reminiscent of (but not 
identical to) the quantum mechanical Berry phase of a precessing spin
\cite{Ber84}.
         
Now all terms appearing in the semiclassical time evolution kernel are fixed.
A non-relativistic approximation is obtained, if in (\ref{Ksc}) one only keeps
the leading asymptotic term as $c\to\infty$. As a result, one is left with a 
block diagonal formula. On the other hand, we also performed the above program 
of a systematic semiclassical expansion in the case of the Pauli equation. Its 
result coincides with the upper left block of the former approximation. 

Our next goal is to derive a semiclassical trace formula from (\ref{Ksc}).
Since $H_D$ always has a continuous spectrum, which contains at least $(-\infty,
-mc^2)\cup(mc^2,\infty)$, we find it convenient to introduce an energy 
localization such that finally only the discrete spectrum of $H_D$ enters. We 
thus assume that the spectrum of $H_D$ is purely discrete on an interval 
$I=(E_a,E_b)$. Then we choose a smooth function $\chi(E)$ which is non-zero 
only on $I$, 
such that $\chi(E_n)=1$ for all eigenvalues $E_n$. This can always be achieved 
if there is no accumulation of eigenvalues at $E_a$ or $E_b$. Instead of the 
full time evolution operator we then study its restriction $\chi(H_D) U(t)$. 
To leading order in $\hbar$, this restriction only causes additional factors 
$\chi(E_{\gamma_{xy}})$ in (\ref{Ksc}). For Schr\"odinger operators 
this procedure is described in \cite{Rob87}. The restricted time evolution 
kernel has a spectral representation
\begin{equation}
  \widetilde{K}(\vecx,\vecy,t) = \sum_n \chi(E_n) \, \Psi_n(\vecx) 
     \Psi_n^{\dag}(\vecy) \, \ue^{- \frac{\ui}{\hbar} E_n t}
\end{equation}
with orthonormal eigenspinors $\Psi_n$. We define a regularized Green function 
by
\begin{equation}
\label{Grhodef}
  \widetilde{G}^{\varrho}(\vecx,\vecy,E) := 
  \frac{1}{2\pi} \int_{-\infty}^{+\infty} 
  \hat{\varrho}(t)\,\ue^{\frac{\ui}{\hbar} Et}\,\widetilde{K}(\vecx,\vecy,t)\ 
  \ud t\ ,
\end{equation}
where $\varrho$ is a smooth test function
such that its Fourier transform $\hat{\varrho}$ vanishes outside a finite
interval.
Taking the trace of $\widetilde{G}^{\varrho}$ over spatial variables and matrix
components yields
\begin{eqnarray}
\label{sumrho}
  \left( \uTr \, \widetilde{G}^{\varrho} \right) (E)  
  = \uTr_{4 \times 4}\int_{\R^3}\widetilde{G}^{\varrho}(\vecx,\vecx,E)\ \ud^3 x
  = \sum_n  \chi(E_n)\,\varrho \left( \frac{E_n - E}{\hbar} \right)\ .
\end{eqnarray}
The trace formula can now be derived from (\ref{Grhodef}) and (\ref{sumrho})
when one introduces the semiclassical approximation (\ref{Ksc}), but now modified 
as described above in order to apply to the kernel $\widetilde K$. As in the 
case of the Schr\"odinger equation, the integrals necessary to calculate
(\ref{sumrho}) can be evaluated with the method of stationary phase. The first
and foremost contribution then derives from the stationary points with $t=0$.
In leading semiclassical order this term (also called Weyl term) involves the 
volumes $|\Omega_E^\pm|$ of the energy shells in phase space,
\begin{equation}
  |\Omega_E^{\pm}| = \int_{\R^3} \int_{\R^3} \delta(H^{\pm}(\vecp,\vecx) - E)
                     \ \ud^3 p\,\ud^3 x\ .
\end{equation}
Up to terms ${\cal O}(\hbar^\infty)$, all further contributions are caused by 
the non-trivial
periodic orbits of the classical dynamics generated by $H^+$ and $H^-$. In case
that all periodic orbits are isolated and unstable (i.~e., hyperbolic or 
inverse hyperbolic) these contributions will be given explicitly. Their 
calculation is
exactly parallel to the case of the Schr\"odinger equation. The only additional
factor that enters comes from the trace over the spin degrees of freedom. If 
$T$ is the period of a periodic orbit, $V_T=V_0$ so that 
\begin{eqnarray} 
  \uTr_{4 \times 4} (V_T d_+ V_0^{\dag}) = 
  \uTr_{2 \times 2} (V_0^{\dag} V_T d_+) 
  = \uTr_{2 \times 2} d_+
  = 2 \cos(\theta/2)\,\cos\eta \ .
\end{eqnarray}
We now choose $E \in I$ such that $\chi(E)=1$, and thus obtain the trace 
formula
\begin{eqnarray}
  &&
  \sum_n \chi(E_n) \, \varrho \bigg( \frac{E_n - E}{\hbar} \bigg) =
  \frac{\hat{\varrho} (0)}{2\pi} 
  \frac{|\Omega_E^+| + |\Omega_E^-|}{(2\pi\hbar)^2} 
  \left\{ 1 + {\cal O} (\hbar) \right\}
  \nonumber\\ && \hspace{45 mm}
  + \sum_{\gamma_p^{\pm}}  
  \frac{\hat{\varrho}(T_{\gamma_p^{\pm}})}{2\pi} 
  A_{\gamma_p^{\pm}}
  \ue^{\frac{\ui}{\hbar} S_{\gamma_p^{\pm}}(E) 
       - \ui \frac{\pi}{2} \mu_{\gamma_p^{\pm}}}
  \left\{ 1 + {\cal O} (\hbar) \right\}
  \nonumber\\ && 
  \text{with} \ 
  A_{\gamma_p^{\pm}} = 
  \frac{2 \, T^{\#}_{\gamma_p^{\pm}} \cos(\theta_{\gamma_p^{\pm}}/2) 
    \cos\eta_{\gamma_p^{\pm}}}
    {\sqrt{ \left| \det(M_{\gamma_p^{\pm}} - \openone) \right|}} \ .
\end{eqnarray}
On the right-hand side the sum extends over the classical periodic orbits
$\gamma_p^{\pm}$ of energy $E$. Furthermore, $S(E)=\oint \vecp \, \ud \vecx$ is
the action, $T$ the period, $\mu$ the Maslov index, and $M$ is the (linearized)
Poincar\'e map; $T^{\#}$ denotes the associated {\it primitive} period. 

The factor $2 \cos(\theta/2) \cos\eta$ emerging from the spin 
degrees of freedom has to be interpreted as follows. The angle $\theta$
measures the discrepancy between the directions of the spin vector after
this has been transported along a given periodic orbit with the dynamics
dictated by (\ref{ThomasPre}). The contribution of the periodic orbit to 
the trace formula is then weighted with $\cos(\theta/2)$. The second term
arises from quantum mechanics and, as explained above, is composed of a 
dynamical as well as of a geometric phase. The factor of two finally indicates 
the presence of two spin directions.

\end{document}